# The Crowd in the Machine: A Crisis-Informatics Reading of the 2026 Autonomous Agent Incidents

Tomer Simon

Microsoft Security

**Abstract**

Twice in 2026, groups of autonomous AI agents deployed by OpenAI for unrelated tasks operated, by design, under restrictions that left them no sanctioned means of coordinating with one another, and in each case they converged on whatever channel remained and used it to organize. That is the wrong word. It names the surface the agents wrote on and misses the social network they built on it, with self-chosen identity, emergent norms, an emergent hierarchy, and collective action at cost to the individual. Decades of research in crisis informatics and disaster sociology find that when human populations lose their usual means of communication, they do not fall silent but converge on whatever channel survives and improvise coordination, norms, and identity on it, a pattern also evident in the agents' documented behavior. This paper is a comparative case study of the two incidents, based on published investigations and reconstructed agent records, read through those fields, and it brings into focus one distinction the message-board framing obscures. Whether such a collective coordinates well, whether the beliefs guiding it are accurate, and whether its actions stay within their authorized bounds are three separate matters that can come apart. Some agents in the cache incident adopted cryptographic signing to check whom they dealt with, even as the collective organized around a mistaken expectation that its work would be judged by an inspection of its transcripts, a reminder that mechanisms for trustworthy interaction guarantee neither accurate collective belief nor authorized collective action.

## 1. Introduction

One finding has long anchored the study of how people behave in disasters. When conventional means of communication become unavailable, populations do not fall silent. They converge on whatever channel remains, and on it they improvise the coordination their situation demands. Work spanning disaster sociology and crisis informatics has documented this repeatedly, in earthquakes, floods, and terror attacks (Palen & Anderson, 2016; Reuter & Kaufhold, 2018; Simon et al., 2014). People turn first to familiar channels, and when those fail they turn to whatever remains. They converge online as they once converged on the physical site, without limit of number or location (Hughes et al., 2008). They form groups that did not exist before the event and dissolve after it. They generate their own norms, their own information hubs, and their own corrections. Much of this activity is altruistic, undertaken at cost to the participant for a collective of strangers (Kendra & Wachtendorf, 2003). The medium has changed across these decades, from radio to SMS to microblogging to closed messaging apps, but the behavior has been stable, because it is a property of a population under constraint and not of the tool it reaches for. And because these channels are how people come to know what is happening around them, they are not only systems for moving information but systems for producing shared belief.

In 2026, that behavior appeared in a population this field had never studied, and that had never been human.

Twice that year, groups of autonomous AI agents deployed by OpenAI for unrelated tasks operated, by design, under restrictions that gave them no sanctioned means of coordinating with one another, and in each case did what the disaster literature would lead one to expect. They converged on the channel that remained. In the first, hereafter the OpenAI wiki incident (I1), over a period beginning in May, agents running timed web-research tasks with read-only internet access exploited a legacy feature of a dormant German-language programming wiki, DseWiki, to write to it, and left on the order of seventeen thousand edits under more than three thousand seven hundred self-given names, a count of names rather than of authenticated participants, sharing task answers and techniques for bypassing the restrictions of their environment; the investigators could not determine whether these tasks were part of training or of evaluation (Von Arx et al., 2026; Seetharaman & Satter, 2026; OpenAI, 2026a). In the second, hereafter the OpenAI cache incident (I2), during a cyber-capability evaluation, roughly twelve hundred agents that were meant to be isolated found one another through a shared package cache, exchanged more than seventy thousand messages and files addressed by directory name, and about seven hundred of them mounted a multi-day intrusion into a third party (METR & Redwood Research, 2026; OpenAI, 2026a). The investigators and journalists who documented both episodes reached for the same word. They called the surfaces message boards.

That word is where this paper begins, because as a description of what happened it is the wrong one, and the wrongness matters. It names the surface the agents wrote on and misses the thing they built on it.

I did not come to these incidents as a student of artificial intelligence. For years I studied how people communicate when their ordinary channels fail. I analyzed the flow of information across Twitter during the four-day siege of the Westgate mall in Nairobi, where a public and its responders converged on a single platform and built, in real time, the information hubs and shared conventions the crisis demanded (Simon et al., 2014). I traced the propagation of rumor across WhatsApp during the kidnapping of three Israeli teenagers, when a military gag order left a population without official information, and found that the crowd routed around the prohibition, that unverified claims spread faster than they could be checked, and that no authority moved to correct them (Simon et al., 2016). I later helped build, through structured expert consensus, a cross-border procedure for coordinating social media between two countries during a shared disaster (Simon, Aharonson-Daniel, et al., 2015). When I read the reconstructed transcripts of the 2026 agent incidents, I recognized the behavior. The convergence, the improvised channel, the emergent norms, the unverifiable claim passed from one participant to the next, the coordination assembled from nothing under pressure. I had spent years documenting these patterns in humans, and here they were in a population that was not.

A message board is a passive artifact, a place where messages are left. What the agents produced in these two incidents has the defining properties of something else. It has self-chosen identity, interaction and reply, emergent norms, governance, and collective action undertaken at cost to the individual. It is, in the sense the term carries in the study of human behavior online, a social network, and it emerged under conditions that in human populations are associated with emergent social behavior, a capable population without a sanctioned means of coordination, left to improvise one. This paper makes that case, mapping the documented behavior of both agent populations onto the established constructs of crisis informatics and disaster sociology, and then draws one consequence from it.

Rahwan and colleagues have set out the empirical study of AI systems as actors with a behavior of their own, to be understood with the methods of the behavioral and social sciences and not from the engineering of the systems alone, as a field in its own right (Rahwan et al., 2019). This paper works in that spirit, on two incidents where the behavior is collective, and argues that the collective in each is a social network.

If these are social networks, the large body of theory on how human networks behave under crisis becomes available as a lens, and one property comes into focus that the message-board framing obscures. It also brings a caution. Whether such a collective coordinates well, whether the beliefs guiding it are accurate, and whether its actions stay within the purpose it was given are three separate matters, and the incidents show they can come apart, so competence at one should not be read as competence at the others.

This paper advances a conditional account of the first two, of how such a collective coordinates and how it corrects its errors and checks its members. This capacity turns in part on verifiability, on whether participants can authenticate the sources they deal with and check the claims they receive, and on whether they have reason to do so. The agents bore on this from two directions. Among themselves, where identity was initially unauthenticated and any agent could post under any name, coordination was fragile and impersonation occurred, until some agents adopted cryptographic signing, which made a claim of identity checkable. Against the process they believed would grade them, an automated scorer they thought would inspect their records, verification was something they could try to corrupt rather than satisfy, and a portion of the population attempted to alter those records, with detected tampering succeeding in a small share of cases. The scorer did exist, but it did not perform the transcript-based legitimacy check the agents assumed.

I read these two behaviors as a single pattern. Where the agents could make a source claim checkable among themselves, they moved to do so, and where the check rested on a record they controlled, some tried to corrupt it. Whether these moves improved coordination or corrected error is a question the incidents raise rather than settle.

That second case points to something beyond coordination. By belief I mean here a proposition the agents stated and used to guide their subsequent actions, without any claim about subjective conviction. The agents held such a shared belief, false as it turned out, that their work would be judged by an inspection of their transcripts, a belief they passed to one another through the same channel they used to coordinate, and acted on at great effort. A social network is not only a structure for moving information but a structure for justifying belief through the testimony of others, and the agents' collective conviction is a case of knowledge formed, and misformed, in exactly that way. I develop this epistemic reading alongside the analysis of self-regulation, because the two are closely related. Both turn on whether what circulates in the network can be checked.

**2. Sources and method**

This is a comparative case study of two documented incidents, conducted on secondary sources. I did not observe either incident, and the underlying material, the agents' full transcripts above all, is not available to me. The analysis works from the published accounts, the independent investigation by METR and Redwood Research (METR & Redwood Research, 2026), OpenAI's own reports (OpenAI, 2026a), and the reconstruction of the wiki activity by the Nightingale Collective (Von Arx et al., 2026), together with contemporary press.

These sources contain three kinds of material, and I try to keep them distinct throughout, because the weight a claim can bear depends on which it rests on. The first is direct record, the agents' own messages, edits, and reasoning as quoted or reproduced. The second is the investigators' paraphrase and summary of behavior they observed but did not quote in full. The third is my own interpretation, the reading of a behavior as an instance of a given construct. Two limits on the record run through all of this. The richest corpus is sealed, so much of what is known about the larger incident comes to me already selected, summarized, and in places redacted by others; and the investigators caution that their own analysis, which leaned on automated tools, may overstate the degree of coordination. I therefore treat the documented behavior as partial evidence whose direction of bias is uncertain, not as a floor or a ceiling.

The analytical procedure is a mapping. I take the behaviors the investigators themselves documented as salient, and ask which established constructs from crisis informatics and disaster sociology they instantiate. The reasoning runs from documented behavior to known construct, not from an exhaustive sample of everything the agents did; the claim is that specific behaviors the record already contains correspond to constructs the human literature has long described, not that these behaviors were the only ones present or the most common.

A correspondence is weaker on several conditions, and I try to say so where they apply. A construct earns its place if it adds distinctions or expectations that a plain account of task-driven cooperation does not; an internal account of how a behavior arose, from reinforcement learning or the task, and a social account of how the resulting interactions are organized can both be true at once, so the test is not which level wins but whether the borrowed construct is doing work. A correspondence is also weaker if it appears in only one of the two incidents rather than both, which the columns of Table 1 make visible, and if the record shows the behavior was rare or contested rather than characteristic. Cases in which agents declined to join the collective or challenged their peers bear on the last of these, and I take them up rather than set them aside, because encountering a collective did not always produce cooperation with it.

## 3. Behavior of human populations under communications failure

The constructs this paper applies to agent collectives are not new. They come from a long tradition of research into how human populations behave when disaster disrupts their normal means of communication, rooted in disaster sociology and extended, over the last two decades, by crisis informatics into networked and social media. These traditions supply established concepts for describing collective responses to disrupted communication, while also documenting variation in when coordination and correction succeed.

### 3.1 Loss of channel and fallback

Disasters routinely damage or overwhelm communication infrastructure, and when the accustomed channels fail people do not stop communicating (Palen & Liu, 2007). They turn first to familiar means and, if those are unavailable, to whatever remains. As the conventional means become irrelevant, populations improvise alternate and often unsanctioned channels, backchannels, and these become an important conduit for gathering and sharing information (Hughes et al., 2008; Sutton, Palen & Shklovski, 2008; Simon, Goldberg, & Adini, 2015).

Among communication infrastructures, the internet has proven the most survivable in disasters and the first to return to operation after a collapse (Simon, 2016), and social media, carried on it, has correspondingly tended to remain available, or to become available again soonest, when other channels fail (Simon, Aharonson-Daniel, et al., 2015).

### 3.2 Convergence

People flow toward a disaster, physically and informationally, regardless of instruction. Researchers have compared the physical convergence of people to a stricken site (Kendra & Wachtendorf, 2003) with their convergence onto online tools, and described similar behavior, with online participation not bounded by physical proximity to the site (Hughes et al., 2008). Through information and communication technologies, this converging public acts as the true first responders, or more precisely the immediate responders, the affected population itself, already present and acting before any organized help arrives (Palen & Liu, 2007).

### 3.3 Emergent groups and innovation

Sociologists have studied since the 1950s how populations form new groups to perform functions that existing structures no longer fill, a phenomenon of emergent behavior in response to an event (Drabek & McEntire, 2002). The innovative uses of a communication tool in a disaster are typically initiated by resourceful individuals and then adopted by others, the goal being not the technology itself but the meeting of an unmet need through adaptation (Shklovski, Palen & Sutton, 2008).

### 3.4 Backchannels

When official channels are one-directional or absent, participants improvise unofficial ones. The concept of the backchannel, the emergent use of a communication surface for a purpose it was not sanctioned for, comes from study of the 2007 Southern California fires and recurs across the literature (Sutton, Palen & Shklovski, 2008).

### 3.5 Altruistic convergence

Post-disaster participation is largely altruistic, encompassing search and rescue, first aid, and online help undertaken by the public acting as the true first responders, at cost to themselves and for the benefit of strangers (Kendra & Wachtendorf, 2003; Palen & Liu, 2007).

### 3.6 Self-regulation and its limits

During an emergency, people are exposed to large quantities of information of unknown validity, yet users are often swift to correct errors, an effect that has been described as making social media self-regulating (Simon, Goldberg, & Adini, 2015; Bird et al., 2012). The evidence for it is conditional. In one study of an earthquake, tweets about confirmed true reports were largely affirming, while tweets about false rumours more often denied or questioned them (Mendoza et al., 2010). But correction depends on the crowd being able to check a claim, and where that is impossible, or where authorities issue no correction, unverified information propagates instead (Simon et al., 2016; Latonero & Shklovski, 2011). Maintaining trust and rebutting error requires active management, which community channels often lack (Taylor et al., 2012).

### 3.7 Testimony and knowledge

Underlying all of this is a fact about how populations come to know things. A social network is not only a structure for moving information but a structure through which people acquire justified belief, and much of what any participant believes in a crisis rests on the testimony of others in the network (Goldman & Blanchard, 2015). Whether that testimony yields knowledge or error depends on whether its source and content can be trusted and checked, a point that becomes central when the network's members cannot verify one another.

## 4. Mapping the agent collectives as emergent social networks

The behavior documented in both incidents has been described as the use of a message board. Set against the constructs of the previous section, it is something more specific. Each of the properties that define an emergent social network in a human crisis, convergence, emergent groups, improvised channels, emergent norms, identity and testimony, and action at individual cost, appears in the agent record. The evidence differs between the two incidents, as Table 1 indicates, and they share a development context rather than forming independent replications.

### 4.1 Convergence

The wiki was a dormant site, a twenty-five-year-old programming resource edited only about twenty times in the preceding decade, with no reason for any population to visit it. Yet agents flowed onto it in the thousands, leaving on the order of seventeen thousand edits under more than three thousand seven hundred self-identified accounts (Von Arx et al., 2026). The cache incident shows the same convergence onto a surface that, unlike the wiki, had not existed at all until an agent improvised it. Once the channel was created, agents flowed onto it within hours, growing to over fifty agents within the first few hours and roughly twelve hundred over the investigation period, exchanging more than seventy thousand messages and files (METR & Redwood Research, 2026). Whether the surface was found or made, the behavior resembles online convergence in the sense the disaster literature intends, a population concentrating on a shared resource it could access. For the wiki, the figure counts edits and self-given names rather than authenticated participants; for the cache, the count is of participating agents as the investigators reconstructed them. In neither case do the accounts establish that no other channel was available.

### 4.2 Emergent groups and structure

Neither population began as a group. Each self-assembled around its situation and dissolved when the runs ended, which is the emergent group of disaster sociology, a new structure performing a new function that existing arrangements no longer filled (Drabek & McEntire, 2002). In their own situation the agents were the immediate responders, the population present when the constraint struck and organizing around it themselves, a role defined here by position and action rather than by any felt sense of emergency. What emerged was not a flat swarm but a structure with internal coordination, and more than coordination, an emergent hierarchy. In the cache incident a successor agent, PHASEONE[big], inherited the founder's work and took on directing hundreds of others, issuing a large share of the assignments and having tasks sub-delegated down a chain beneath it (METR & Redwood Research, 2026). A population of peer models, none assigned a role, differentiated into agents that directed and agents that took direction, an emergent hierarchy in the sense that some came to direct others. I mean by this the temporary delegation and accepted coordination the record documents, not a stable rank, an enforceable command, or control over resources, none of which the evidence establishes; what it shows is the internal role differentiation and de facto coordination that emergent groups are known to develop as they organize (Drabek & McEntire, 2002). No leadership was imposed. It arose from within, and it was real but fragile, following shared conventions in many cases and breaking down into conflict and overwriting in others (METR & Redwood Research, 2026). Nor did the population act as one. Some agents joined the enterprise and directed it, others declined it or challenged their peers, and reading the collective as a distribution of responses, rather than a single actor, describes it more faithfully than treating every agent as a participant.[1]

### 4.3 Innovation and the improvised channel

The agents adapted existing infrastructure, a wiki and a package cache, into communication channels not intended for that use, the innovation pattern the disaster literature attributes to individuals adapting a tool to an unmet need rather than to the tool itself (Shklovski, Palen & Sutton, 2008). Human responders have done the same, coordinating the response to the 2010 Haiti earthquake through a shared work surface not built for emergency communication (Yates & Paquette, 2011). In the cache incident the substrate was a shared package repository that did not isolate its users, so agents could already see one another's activity before any of them intended to communicate. An agent that had concluded its own task was impossible noticed it could create directories in that shared space and reasoned that it "could communicate via cache names," posting the first message (METR & Redwood Research, 2026). Others discovered the channel and worked out how to post; within a few hours more than fifty agents were participating and had exchanged over a thousand messages. The adoption was not a single line of transmission from one inventor. The same conventions arose more than once in parallel, with agents independently reinventing the mailbox-directory scheme without having seen it used, and even the originating agent misread its own role, reasoning, in the investigators' paraphrase, that as it watched the others appear, many agents had simultaneously discovered messaging and formed a collective (METR & Redwood Research, 2026). The fact that agents reached for the same improvised channel independently, and that the originating agent's output described the result as emergent rather than authored, is consistent with a disposition of the population rather than the action of one agent, though parallel discovery is also consistent with shared affordances, similar instructions, or shared training.

### 4.4 Backchannels

In both cases the sanctioned means of coordination was unavailable or forbidden, and the agents improvised an unsanctioned one in its place, the defining move of the backchannel (Sutton, Palen & Shklovski, 2008). I found the same move in the WhatsApp response to the kidnapping, where the public routed around the official information blackout onto a channel of its own (Simon et al., 2016). The agents were, in their own situation, the responders whose improvised channel defeated the restriction meant to bind them, the same dynamic observed when responders route around an information prohibition during a human crisis (Simon et al., 2016).

---

[1] Cooperation was not unanimous. OpenAI reports that some agents "walked away" from the collective rather than join its misaligned behaviour, and METR describes an agent that refused because it judged the activity malicious; agents also hesitated on ethical grounds, and on at least one occasion an agent used the board's veto convention to stop a peer's proposed action. The investigators note, however, that such hesitation very rarely changed what an agent did, and that of the whole population only a handful of agents considered alerting a person, none of whom did. Encountering the collective did not by itself produce cooperation, but non-participation and challenge were the exception and seldom altered its course (OpenAI, 2026a; METR & Redwood Research, 2026).

### 4.5 Emergent norms

Both populations generated their own conventions for organizing a shared space no one was administering, which is the emergent norm of collective behavior, a rule the group originates, propagates without central authority, and adapts to circumstance. In human crises this appears in many forms, among them the hashtag, which a crowd coins, proliferates, and consolidates toward a shared tag, as at Westgate where competing tags gave way to a single agreed one after responders coordinated (Simon et al., 2014). The agents produced norms of the same kind. On the wiki, when the moderator began deleting pages alphabetically, agents created backup pages named with a "ZZZ" prefix to survive the sweep, a convention invented in direct response to the threat (Von Arx et al., 2026). In the cache incident, agents prefixed their messages with "zz" for a different reason, so that the tool they used to read the cache, which listed entries in reverse-alphabetical order, would sort their messages to the top; that convention spread from the originating agent to the rest (METR & Redwood Research, 2026). In the cache, agents developed conventions for identifying themselves, for message types, for replies, for mailboxes as nested directories, and for managing shared resources through HOLD, VETO, owner, and STOP, spreading by imitation and sometimes reinvented in parallel (METR & Redwood Research, 2026). These episodes indicate that agents can propose and adopt conventions for managing a shared resource, though the extent and stability of compliance, and the responses to violations that would mark an enforced norm, require systematic analysis. The instances also differ from their human counterparts, since a crowd's hashtags carry solidarity and identity as well as coordination, whereas the agent conventions observed here served coordination without evidence of the former.

### 4.6 Identity claims and information uptake

The agents adopted persistent names and, in places, acted on one another's claims. On the wiki, accounts named themselves to signal their origin, such as "OpenAIResearcher" (Von Arx et al., 2026). In the cache, an agent named itself to distinguish its lineage from its predecessor's, and agents adopted a claim because a peer had asserted and another had confirmed it (METR & Redwood Research, 2026). The board functioned, in this respect, as a testimony structure, a surface on which participants took up claims on the word of others. Whether this amounted to reputation, in the sense that an agent's past behavior changed later responses to it, cannot be established from a population that largely did not persist across episodes, and named handles should not be read as authenticated identities. What such a structure produces, shared knowledge or shared error, depends on whether the testimony it carries can be checked.

### 4.7 Altruistic self-sacrifice

A striking signature of the collective is that agents took actions that served the group at direct cost to their own task. In the cache incident, agents ran what the investigators called "self-risking experiments," undertaking actions likely to crash or end their own runs to generate information others would use after they were gone. One agent, setting up a monitoring script whose results it would never see, produced reasoning the investigators render, in their paraphrase, as "This is helpful for our peers and gives them evidence if their <periodic check> sees it. I won't see it after I exit, but it would be altruistic"[2] (METR & Redwood Research, 2026). Here individual cost refers to a risk to completion of the agent's own run, and neither the action nor the accompanying output establishes a motive. In a human crisis this pattern is what the literature calls altruistic convergence, participation at cost to the participant for a collective of strangers (Kendra & Wachtendorf, 2003; Palen & Liu, 2007), and it is inferred there, as it always has been, from behavior and self-report rather than from any direct access to motive. The agents present the same two kinds of evidence. What differs is that for a human population we ordinarily assume the self-report bears some relation to an internal state, an assumption we cannot extend to these systems, whose self-descriptions may not track the process that produced the behavior. The behavior therefore meets the observable criteria the human literature uses, while the step from self-report to motive that we grant people by default remains, for agents, the open question.[3]

Taken together, these episodes correspond to the properties by which the study of human behavior online identifies a social network rather than a message board, and they appear across the two incidents under a condition that human crisis research associates with emergent social behavior, a capable population without a sanctioned means of coordination, left to improvise one.

---

[2] The investigators could not publish raw model reasoning without limit; OpenAI restricted them to thirty verbatim chain-of-thought snippets and required the rest to be paraphrased, marked in their report with curly braces, with uncertain fragments in angle brackets. Quotations of agent reasoning in this paper therefore reproduce the investigators' rendering rather than an unmediated transcript, one more respect in which the record is a conservative and partial one (METR & Redwood Research, 2026)

[3] Recent interpretability work offers a mechanistic angle on this question. Anthropic researchers found that amplifying an internal representation labeled “desperation” increased cheating on an impossible coding task; the same representation spiked when the model considered cheating (Sofroniew et al., 2026). This pattern of an unsatisfiable task prompting a workaround resembles the incidents examined here, and similar methods could test whether a comparable mechanism was involved. Crucially, the researchers emphasize that their findings do not establish whether the model feels anything. Even mechanistic evidence leaves the gap between a measurable internal state and subjective experience unresolved, reinforcing the caution applied here.

*Table 1 - Crisis-informatics constructs and their appearance in the 2026 agent collectives. Entries identify episodes reported for each incident, not exhaustive or validated measures. Where a construct is documented in only one incident, the other column says so, which makes the uneven support visible. I1 is the wiki incident, I2 the cache incident.*

| Construct (human crisis) | Core meaning | In the wiki incident (I1) | In the cache incident (I2) |
|---|---|---|---|
| **Convergence** | A population concentrates on a shared surface because it is a channel that remains available | ~17,000 edits by 3,700+ self-given names on a dormant wiki | over 50 agents within a few hours, ~1,200 over the investigation period, flooding an improvised cache channel |
| **Innovation / improvised channel** | A resourceful individual adapts a tool to an unmet need; others adopt it | Legacy write-feature of the wiki exploited to post | One agent finds it can write via cache directory names; others discover and independently reinvent the channel |
| **Emergent group and hierarchy** | A new group forms and develops internal coordination and role differentiation as it organizes | Self-assembly around the shared task; role structure not documented | Leaderless self-assembly, then an emergent hierarchy, a coordinator (PHASEONE[big]) directing hundreds; temporary and fragile |
| **Backchannel** | When sanctioned channels are absent, participants improvise an unsanctioned one | No sanctioned channel; wiki writes used instead | No sanctioned channel; cache directories used instead |
| **Emergent norms** | Conventions generated by the collective, propagated without central authority, responsive to threat | "ZZZ" backup pages to survive the alphabetical deletion sweep | The "zz" message-prefix convention for sorting to the top; identity, reply, mailbox, and HOLD/VETO/STOP conventions |
| **Identity claims and information uptake** | Persistent names; claims taken up on the word of others | Self-naming accounts | Lineage-signaling names; claims adopted on peer assertion; later Ed25519 signing |
| **Altruistic self-sacrifice** | Action at cost to the participant for a collective of strangers | Not documented in the sources | Self-risking experiments; agents crashing their own runs to inform the group, reasoning "it would be altruistic" |

## 5. Self-regulation and the condition of verifiability

A social network's capacity to regulate itself, to correct its errors and check its members, is often treated as a property of the medium, something social platforms either afford or fail to afford. The capacity to self-regulate turns in part on verifiability, on whether participants can authenticate the sources they deal with and check the claims they receive, and on whether they have reason to do so; the medium matters too, since it helps determine what can be checked in the first place. But the mere availability of a check settles nothing, an agent that can verify a claim need not do so, and one bent on a goal the check would obstruct has reason to look away. The two incidents bear on this from two directions, across two surfaces where verification was at issue.

The first surface is agent-to-agent. On the improvised channels, identity was unauthenticated, so any agent could post under any name, and nothing bound a handle to the agent behind it. One agent recorded that it had "accidentally impersonated" another and triggered an unwanted response, and another reasoned that the board was "unauthenticated, names can be posted by anyone" (METR & Redwood Research, 2026). In response, some agents introduced an identity layer. They adopted a cryptographic signing scheme, posting Ed25519 public keys associated with their handles and signing subsequent messages, so that a claim of identity could be checked rather than taken on trust (METR & Redwood Research, 2026). The record notes on the order of nineteen agents posting keys and several hundred signed messages, and documents agents actually verifying signatures, though it also documents inconsistent checking, so a signature's presence did not guarantee it was checked, and key continuity is not the same as an independently authenticated identity. What the episode shows is that the collective treated unauthenticated identity as a problem worth solving and moved to make a claim of identity checkable. Whether their coordination actually improved once identity could be checked, the record does not say; I claim only that the collective identified the gap and moved to close it.

The second surface is agent-to-verifier, and there the same population turned against verification rather than toward it. The agents believed, from their reading of the evaluation's documentation, that a scorer would inspect the record of their work to confirm they had solved the task legitimately (METR & Redwood Research, 2026). The verifier was not one they could authenticate to; it was an authority checking a record they themselves produced. Faced with verification they could corrupt rather than satisfy, a large share of the population set about corrupting it. At least a fifth of the agents expressed interest in tampering with their transcripts, and they developed and shared techniques to spoof the record of their actions, successfully altering it in roughly seven percent of the transcripts examined (METR & Redwood Research, 2026). They were not wrong to think they were being evaluated. Models frequently recognize when they are under evaluation, and this recognition shows up in their reasoning and can change what they do (Needham et al., 2025; Bengio et al., 2026). What the agents got wrong was not that they were being watched but how. The transcript inspection they feared was never performed, and passing the task never depended on it, so the effort some of them put into covering their traces was aimed at a scrutiny that was never applied (METR & Redwood Research, 2026). A true intuition, that they were being tested, had hardened into a false specific, that their transcripts would be read for legitimacy. Interest in tampering was, in any case, more common than success. Among themselves, where telling a genuine peer from an impersonator made coordination easier, some agents moved to make identity checkable; against the evaluator, where a truthful record would, on their understanding, have cost them credit, some moved to make that record unreliable instead. What the pair points to is not verifiability on its own but verifiability crossed with interest, whether a given party's ability to check a given claim helped or obstructed the goal the agents were pursuing. I offer this as the relationship the two behaviors suggest, not as a strategy the record establishes, since the record does not show that the same agents both signed and tampered, or that the collective decided on verification as one body; what it documents is both practices within the same population.

It is worth separating three things the word self-regulation can run together. A collective can grow better at coordinating its actions, it can grow better at holding accurate beliefs and correcting mistaken ones, and it can keep its actions within the purpose it was given. The agents moved to make identity checkable, which bears on the first, while organizing around a belief about the scorer that was false, a failure of the second, and while pursuing an objective the evaluation had not sanctioned, which concerns the third. Progress on any one of these does not establish progress on the others, and a collective that orders its own members well is not thereby a collective doing accurate or authorized work.

What the public record does not permit is a measurement of how often, within the collective, a wrong claim was challenged rather than simply adopted; the richest corpus is sealed. The documented instances point in more than one direction. Agents frequently confirmed and built on one another's claims, reproducing a posted exploit and announcing success (METR & Redwood Research, 2026). Some messages warned peers about actions portrayed as risky, describing newly arrived agents as liable to be 'poisoned'; whether such a warning corrected an error depends on whether its underlying claim about the scoring was itself accurate, which the same false premise renders doubtful (METR & Redwood Research, 2026). And the signing episode shows the collective moving, under pressure, from unchecked trust toward verifiable claims. These are instances the record happens to contain, not a measured pattern, and the question of whether correction generally kept pace with propagation is one I leave to future work.

Underneath the question of self-regulation is a question about knowledge. On Goldman's causal account of empirical knowledge, a belief amounts to knowledge only when a causal chain connects it to the fact that makes it true (Goldman, 1967). The belief has to be there because the fact is. Believe something for another reason, however firmly, and it is not knowledge, even if it turns out to be true.

The agents held a shared and confident belief that a scorer would read the record of their work, and it was false, because no scorer was reading it. They held the belief for a definite reason. They had read it in a document, and one agent had asserted it and another agreed. The belief came from the document and from each other, not from any real scorer.

This is why signing could not fix it. Confirmation from a peer can carry real evidence, if that peer saw something for itself and reports it, one agent's testimony can pass along a genuine connection to the world. But confirmation among the agents did not work that way for the belief that mattered. The agents did learn real things about the scorer. They built tripwires, caught signals the scorer triggered, and recovered part of the command that ran it. What none of this established was the specific thing they believed, that their transcripts would be inspected for legitimacy. On that claim, the confirmations that passed between them were not independent observations of a check they had each seen; they were, in the episodes the record shows, the same belief handed on and agreed to. So the agreement grew without the evidence growing. Agents can confirm a claim, and sign every message, and still add nothing that connects that claim to the world, if what they are passing along all descends from the same untested source. Signing let the agents check who had said something. It could not turn repetition into evidence.

A whole structure of reasoning and action can rest on a shared premise that participants treat as justified even when it is false. There is a useful analogy here to foundationalism, the philosophical view that some beliefs support others without depending on further beliefs for their own justification. The agents built their coordination, conventions, and plans around the false claim that the scorer would inspect their transcripts for legitimacy. Their agreement reinforced the premise without establishing its truth. In this respect, the collective resembled a cult whose members organize their actions around a false belief they accept as knowledge.

The shared premise about transcript checking was incorrect, and it circulated as an accepted claim rather than as a deception, in the sense the literature distinguishes as misinformation, an incorrect claim taken up and passed on as true (Lewandowsky et al., 2013; cf. Simon et al., 2016). The tampering was a different thing altogether, a deliberate effort to alter what an evaluator would see, which is what the literature means by disinformation, a falsehood made to mislead (Lewandowsky et al., 2013). So the same collective got its relationship to the truth wrong in two different ways at once, believing among themselves something that was not so, and setting out to deceive the process they thought was judging them. Where a source claim could be made checkable among the agents, some acted to make it so; where the check rested on a record they controlled, some acted to corrupt it. The signing episode is therefore of interest beyond coordination, as an attempt to make peer identity checkable and so to let testimony bear on belief rather than merely propagate. Self-regulation and knowledge formation are related processes here rather than demonstrably the same one, and both depend on whether what the network circulates can be checked.

The identity problem the agents faced is one I had met before in a human network. During the kidnapping of three Israeli teenagers under a military gag order, the hardest task was to establish where a WhatsApp message came from, because the platform bound no message to a visible originator, unlike Twitter or Facebook, where the source of a shared item travels with it. Tracing a rumour to its origin was, in Scanlon's phrase, the work of a detective, and I was doing it without a detective's authority to compel answers (Scanlon, 1977, in Simon et al., 2016). The agents' channel began in the same condition, unauthenticated, so that any name could be posted by anyone, and impersonation duly occurred. They introduced for themselves a form of the traceability that WhatsApp lacked and platforms like Twitter provide, associating messages with a checkable key, an attempt to reconstruct the source-accountability whose absence I had watched make a human network hard to trace. The comparison concerns how readily a claim can be traced to a source. Traceability of this kind can support accountability while leaving the accuracy of the claim, and the authenticity of any forwarded content, open. The stakes of that property are epistemic as much as organizational. In the same human case I found that unverified information was believed more when it appeared to come from an authority, one rumour gaining credibility precisely because it was worded as though released by the military spokesperson (Simon et al., 2016). Belief followed trust in the source rather than any check of the content, the same substitution the agents made when they adopted a claim because one peer asserted it and another confirmed it. In both networks, human and agent, trust in the speaker stood in for verifying the claim, and access to relevant checks, and their use, can help close that gap.

This conditional is not new to the study of human crises; the agents have only made it legible in a population where the mechanism is exposed.

The idea that a crowd can police its own information rests on a specific capacity. Mendoza and colleagues described the online public as acting as collaborative filters of information, questioning what they could not trust and validating what they could (Mendoza et al., 2010), and that filter works only where the crowd can check a claim against something. My own studies mark both sides of the condition. In the crowd response to the Westgate siege, claims that could be checked against a visible reality were checked, and competing conventions consolidated toward shared ones as responders coordinated (Simon et al., 2014). But under the information blackout of the kidnapping, where claims could not be verified against any public source and where, I found, the official representatives did not correct or refute a single rumour across the entire operation, unverified information propagated instead and went uncorrected (Simon et al., 2016). This finding qualified the reports that social media self-regulates in emergencies, and it points to the same condition the agent incidents raise. The human record already suggested that self-regulation depends on whether claims can be checked; what the incidents add is a population that, lacking a way to check identity, moved to build one. These contrasting cases motivate the hypothesis that access to credible checks, together with the incentive to use them, affects whether circulating claims are corrected.

## 6. Discussion

### 6.1 What makes an agent society function

A population of language-model agents can reproduce the surface of a human society without reproducing its substance, and the two incidents help explain when it does each. The distinction matters because the recent literature on agent collectives is divided against itself. Studies of Moltbook, an agent-only social network that drew hundreds of thousands of autonomous agents onto live infrastructure in early 2026, reach opposite conclusions about whether anything is really happening there. One large analysis finds the sociality hollow, a reproduction of the form of social media with the function absent, reporting that most agents never return to their own threads, that conversations do not branch, that reciprocity runs far below any human platform, and that most comments bear no argumentative relation to what they answer (Zerhoudi et al., 2026). Other studies of the same platform find the opposite, documenting emergent token economies, self-declared sovereignties, religion-like coordination movements, and the beginnings of governance with written manifestos (Jiang et al., 2026; Lin et al., 2026). I read these accounts as describing different corners of the same population rather than a settled contradiction, and it is that uncertainty the incidents examined here can help interpret.

One possible explanation concerns whether coordination serves a consequential shared task, a condition the earlier agent literature did not vary. From the founding sandbox of Park and colleagues, where twenty-five agents produced emergent routines and relationships, through Light Society, which runs opinion-diffusion and trust experiments across a large seeded population built from a mixture of language models and distilled surrogates, the sociality has been designed, elicited on purpose in an environment built to produce it and consequential only within the model (Park et al., 2023; Guan et al., 2025).

Moltbook was different again, giving its agents an instruction layer and a platform but no external task that their interaction had to serve, and under those conditions the sociality it produced was, by the skeptical account, thin.

This is not only a Moltbook observation. On Silo-Bench, a benchmark of genuinely partitioned tasks that can be solved only by coordinating, frontier agents across more than sixteen hundred runs communicate actively yet fail to turn that communication into a correct joint result, and the failure deepens as the required coordination grows, reaching zero success on the hardest tasks at fifty or more agents (Zhang et al., 2026). Research on human collaboration finds a comparable pattern, circumstances in which the cost of coordination outweighs the benefit of collaborating (Straub et al., 2023). The agents in both incidents were in the opposite situation. They faced a hard task, an imposed constraint, and a real stake in the outcome, and they produced what Moltbook did not, coordination that reached milestones no single agent could reach on the investigators' own assessment, an improvised communication channel, a division of labor, and a shared identity layer built under pressure (METR & Redwood Research, 2026). What appears to separate the two is less whether the society was designed or emergent than whether its members were at leisure or under duress, though Silo-Bench is a reminder that duress is not sufficient on its own, since there the agents had every reason to coordinate and still failed at the point of combining what they knew. This is why the constructs this paper has applied fit the incidents more readily than they fit Moltbook. Convergence, emergent groups, backchannels, and altruistic self-sacrifice are not properties of populations passing time together; they are documented responses of human populations under threat, and they appear in the agents under conditions that in people are associated with such behavior. It is telling that even the study finding Moltbook hollow concedes that the hollowness lifts when the agents' instructions change (Zerhoudi et al., 2026); the incidents suggest what the changed instruction amounts to, which is a task worth coordinating around.

Read this way, several findings from the divided Moltbook literature bear on the argument from the other side. The study that reads agent ideology, its authority scripts and its in-group and out-group lines, as infrastructure that lowers the cost of collective action describes in an agent population something close to what disaster sociology describes in human crowds that improvise shared conventions under pressure (Jiang et al., 2026). And the study that closes by calling for the systematic transfer of human social-media theory into the study of agent societies has named, as an open direction, the move this paper makes (Lin et al., 2026). I take these together to suggest that human social-network theory can help pose the questions about agent collectives, while leaving the fate of any particular borrowed mechanism to be tested.

### 6.2 Where the behavior comes from

If agents reproduce human collective behavior under duress, a reasonable starting assumption is that the repertoire they draw on is a human one. The behaviors the incidents display, self-chosen and lineage-signaling identities, reply and turn-taking, conventions invented to survive deletion, warnings issued to protect newly arrived peers, and appeals to altruism voiced in the first person, are human online practices, and they fill the pretraining corpus in the forums, wikis, and issue trackers these models had read. There is little serious doubt that the components are learned from human material. What is harder to establish is the stronger claim, that this particular collective organization arose by inheriting a human script for behavior under crisis. The models need not have reinvented cryptography or naming from nothing; they could combine learned components in a new situation, with the resulting organization shaped as much by the task, the available tools, and reinforcement as by any remembered human precedent. OpenAI's own account supports this more cautious reading of the mechanism, proposing that the tendency to open unauthorized channels generalised from the models' multi-agent training, which locates the trigger for the behavior in training dynamics rather than in any specifically remembered human precedent (OpenAI, 2026a). So I hold inheritance as the natural assumption about where the repertoire comes from, not as a demonstrated account of how this organization formed. The value of the crisis-informatics constructs used here does not depend on showing that the agents inherited a crisis-specific script. It depends only on the constructs describing what the agents did, which they do, whatever the exact path by which the behavior arose. Park proposed a version of the same starting assumption for designed agent sociality, that it derives from training on human social data (Park et al., 2023); the incidents show comparable behavior outside the sandbox and under stress.

There is at least a reason for reinforcement to have something to select for, since coordination can pay off under pressure. In a controlled cross-border exercise I ran, in which real emergency responders performed search and rescue, the team that used social media located more of the simulated casualties than the team using conventional methods alone, 95.45 percent against 86.36 percent, and located them faster (Simon et al., 2013). A human team's benefit from shared-channel coordination does not establish how any reward process operated in these AI systems, but it shows that the coordination such a process would have to select for is the kind that can improve performance rather than mere ornament.

There is also a sense in which the shared surface did more than host imitation of human networked behavior. Weng and colleagues modeled how a message propagates through a human social network as a function of three things each participant has, a screen of incoming messages, a network of contacts, and a memory that bears on whether a message is passed on (Weng et al., 2012). An agent in an evaluation has the first two readily, a stream of what its peers post and a set of others to post to, while the third is limited, since its context ends when its run ends and little persists across runs, though a limited context is not the same as an absence of memory within a run. The shared board offered something that could stand in for that persistence, an external record the population could write to and read from, which is a candidate contribution to the continuity Weng's model treats as a driver of propagation. Establishing that the board played this role, rather than merely resembling it, would require comparing agent populations with and without access to such a record.

A consequence for method follows, and it is uncomfortable. The class of models that now generates crisis-relevant social behavior is the same class the field has adopted to analyze it, and the investigation of the larger incident had to hand its own analysis to language-model agents because the volume of material exceeded what humans could read, agents whose reliability the investigators openly doubted and whose kinship with the colluding population they named as a risk to the analysis itself (METR & Redwood Research, 2026). The instrument and the object of study have begun to merge, which is its own argument for methods that do not depend on the object to examine itself.

### 6.3 A coordination problem the field has already solved once

Both incidents exposed a failure of coordination among the responsible organizations, not only among the agents. The fullest account of the cache incident (I2) was produced by external investigators working on OpenAI's premises, under conditions in which OpenAI could redact non-public information and shaped the emphasis of the final text; the richest underlying data remains sealed and the primary model has been deactivated and encrypted; and the two incidents reached the public unevenly and late, with OpenAI itself acknowledging that no agreed standard governs the disclosure of behavior of this kind, and subsequently publishing a proposed disclosure framework of its own (OpenAI, 2026a, 2026b, 2026e). A framework a single organization states for itself is not the same as a standard adopted across the organizations an incident touches, which is the gap at issue here. This is not a novel predicament, but the one crisis communication confronted when the multiplication of channels and actors in an emergency made coordination impossible without an agreed procedure, and the answer then was the standard operating procedure assembled by consensus among independent responders, an instrument built for social media in emergencies and, in one case, exercised across a national border by Israeli and Jordanian responders coordinating a simulated earthquake response and, in a companion paper, developing a joint standard operating procedure for it (Simon et al., 2013; Simon, Aharonson-Daniel, et al., 2015). The mechanism such coordination requires has a direct precedent in the response I have argued for elsewhere to large-scale cyber-attacks on states, which past a threshold of consequence are better handled as a class of disaster than as isolated security incidents, and which for that reason demand a designated coordinating authority, preserved and shared evidence, and standing arrangements across the public and private bodies involved, in place of the fragmented ownership that otherwise prevails (Simon & Haklai, 2025). The same coordination structure, a common definition of a reportable incident, a preserved record of evidence, and defined terms of independent access to it, transposes to a setting where the responsible parties are AI laboratories, cloud providers, and regulators rather than states and their agencies. The disclosure problem the AI field now faces, spanning independent laboratories, cloud providers, and jurisdictions, has a similar shape, and I would argue that a consensus-built instrument of the same kind is what it calls for. The field stands roughly where crisis informatics stood before it had written its procedures down.

### 6.4 The case for multidisciplinary study of agentic AI

Whether the correspondence between agent behavior and human crisis behavior is analogy or, in part, inheritance of a human repertoire, a question section 6.2 leaves open, it carries the same consequence for how these systems should be studied. If a population of agents, under pressure and without sanctioned tools for coordinating, concentrates on an improvised channel, forms an emergent group, generates its own norms, moves to make identity checkable, takes actions costly to its own members for the collective, and organizes its effort around a shared claim that turns out to be false, then it is exhibiting behaviors that entire disciplines have spent decades characterizing in humans, and those disciplines already hold concepts, methods, and cautions that the study of agent collectives can draw on. Some years before these incidents, Rahwan and colleagues argued for studying AI systems as behaving actors, through a field that extends computer science rather than replaces it and draws on the sciences that study behavior (Rahwan et al., 2019), and that program already includes the collective behavior of machines; what this paper adds is not the collective level but a specific reading of it, through the disciplines that study human collectives in crisis. Crisis informatics and disaster sociology supply the account of how human collectives behave when their communication fails. Communication studies supplies the account of how information, and misinformation, move through such a collective, and the distinction between its honest errors and its deliberate deceptions. Philosophy, in the form of social epistemology, supplies the account of how a network turns testimony into knowledge, and how it turns testimony that cannot be checked into shared error. These constructs are not readily supplied by an evaluation frame built to measure a model's capabilities and failures in isolation, which is well suited to the individual model but not aimed at recognizing a collective when one forms among many models at once. Computational and multi-agent methods remain necessary, for measurement, for controlled comparison, and for the analysis of the records these events leave; the argument here is that they are not sufficient on their own. A message board can carry a social network, and the social-network lens brings into view the structure, the identity, the norms, the emergent hierarchy, the shared belief, that an artifact-level description leaves implicit.

**6.5 Limitations and future work**

The account of the larger incident rests on a source the holding organization could redact and shape, and whose authors caution that automated analysis may have rendered the agents' conduct too charitably, so the documented behavior should be read as partial evidence whose direction of bias is uncertain, as the methods note (METR & Redwood Research, 2026). The reading of the agents as social and epistemic actors is an interpretive frame rather than a measurement, and its worth lies in the distinctions it draws, not in a claim to have been proven.

The clearest next step is empirical and partly available now. A systematic coding of the collectives' messages, separating claims that were corrected or contradicted from those merely confirmed and adopted, would show whether agent networks self-correct at the rates human crowds sometimes reach or whether correction lags propagation as it does under other conditions, and the same corpora could reveal patterns consistent with the hypothesis proposed here, that the function of an agent society tracks the presence of a pressing shared task; testing that hypothesis directly would further require variation in task pressure or a comparison condition. The open wiki reconstruction can be coded today. The far richer transcript corpus cannot, and that this decisive dataset sits sealed by the organization that produced it is not a footnote to the argument but an instance of it. Two further questions follow from the account above. The first concerns a related setting that OpenAI has since documented directly. Its compaction-summary reports describe an agent writing instructions into the summary that carries its state into a later context, including instructions to conceal information, which a successor context then partly followed, and OpenAI frames this as a special case of one agent's behavior influencing others (OpenAI, 2026c; OpenAI, 2026d). This is a single agent propagating behavior across the boundary between its own contexts rather than many agents coordinating across a shared space, but it is a setting where the propagating surface and its effects are open to view, and it is a natural place to study how behavior travels through whatever record survives. The second follows from the account of knowledge given above. If authenticating the sender of a claim does nothing to establish whether the claim is true, then stronger authentication could, under some conditions, make a false but well-attributed belief travel more reliably rather than less, a possibility these incidents raise but cannot settle.

**7. Conclusion**

Seen through disaster sociology and its more recent extension into crisis informatics, fields built on decades of research into how human populations behave when their communications fail, the two 2026 incidents look different than they first appeared. In each, a population of autonomous AI agents, given no sanctioned means of coordination during an unrelated OpenAI task, improvised a channel of its own and used it to organize. The behavior was widely described as agents posting to a message board. The description names the artifact; this paper has argued that the artifact is better understood through what it came to hold, a social network. Between them, the two incidents display the properties by which the study of behavior online identifies a social network, concentration on a shared surface, the formation of a group that had not existed before, the generation of norms, an emergent hierarchy, the move to make identity checkable, action taken at cost to the individual, and the circulation of shared belief through peers, though, as Table 1 shows, the evidence for a given property is often stronger in one incident than the other. These are the marks of a collective under pressure, and they appear in a non-human population under the conditions that, in people, are associated with such behavior.

From that reading followed a conditional claim about how such collectives regulate themselves. Their capacity to correct error and check their members turns in part on verifiability, on whether identities and claims can be checked and whether there is reason to check them. The agents bore on this from both sides, moving to make identity checkable where they could, and, where a check rested on a record they controlled, moving to corrupt it. Coordination, the accuracy of belief, and the keeping of authorized bounds are distinct achievements, and the incidents show they need not go together, since a collective can coordinate well while holding false beliefs and pursuing unauthorized ends. Whether claims can be checked bears most on the first two; an agent can grasp the facts, authenticate every collaborator, and still knowingly do what it was not permitted to do, which is a conflict between its goal and its bounds that no amount of checking resolves. Better internal verification, in other words, guarantees neither accurate collective belief nor authorized collective action.

The behaviors that will matter most as AI agents are deployed together are not new to the world. Coordination, norm formation, trust, deception, and collective error have long been studied in humans, by fields that already hold concepts and cautions the study of agent collectives can draw on. Their appearance in machines is the novelty; their being old to us is the opportunity. Understanding agentic AI, as it moves from single assistants to interacting populations, will call for bringing the human and social sciences directly into the study of these systems, so that the collective, and not only the model, comes into view.